\documentclass[final,3p,times]{elsarticle}
\usepackage{amssymb}
\usepackage[longtable]{multirow}
\usepackage{longtable}
\usepackage{array}
\usepackage{colortbl}
\usepackage{amsmath}
\usepackage{graphicx}
\usepackage[inline]{enumitem}
\usepackage{ragged2e}
\usepackage[hidelinks]{hyperref}
\hypersetup{
colorlinks   = true,
urlcolor     = blue,
linkcolor    = blue,
citecolor   = blue}
\makeatletter
    \g@addto@macro{\UrlBreaks}{\do\/\do\-\do\_} 
\makeatother
\usepackage[ocgcolorlinks]{ocgx2}
\usepackage{multirow}
\usepackage{float}
\usepackage{placeins}
\usepackage{afterpage}
\usepackage{appendix}
\usepackage{enumitem}
\usepackage{subcaption} % Include the subcaption package for subfigures

\journal{Heliyon}

\begin{document}

\begin{frontmatter}

\title{Ergonomic Design of Computer Laboratory Furniture: Mismatch Analysis Utilizing Anthropometric Data of University Students}

\author[inst1]{Anik Kumar Saha\corref{contrib}}
\ead{anikks18@gmail.com}
\author[inst1,inst3]{Md Abrar Jahin\corref{contrib}}
\ead{abrar.jahin.2652@gmail.com}
\author[inst1]{Md. Rafiquzzaman\corref{corauthor}}
\ead{rafiq123@iem.kuet.ac.bd}
\author[inst2,inst3]{and M. F. Mridha\corref{corauthor}}
\ead{firoz.mridha@aiub.edu}

\affiliation[inst1]{organization={Department of Industrial Engineering and Management},%Department and Organization
            addressline={Khulna University of Engineering and Technology (KUET)}, 
            city={Khulna},
            postcode={9203}, 
            country={Bangladesh}}

\affiliation[inst2]{organization={Department of Computer Science},%Department and Organization
            addressline={American International University-Bangladesh (AIUB)}, 
            city={Dhaka},
            postcode={1229}, 
            country={Bangladesh}}

\affiliation[inst3]{organization={Advanced Machine Intelligence Research (AMIR) Lab},
            city={Dhaka},
            postcode={1229}, 
            country={Bangladesh}}
            
\cortext[corauthor]{Corresponding author}
\cortext[contrib]{Authors contributed equally}

\begin{abstract}
Many studies have shown that ergonomically designed furniture improves productivity and well-being. As computers have become a part of students' academic lives, they will continue to grow in the future. We propose anthropometric-based furniture dimensions that are suitable for university students to improve computer laboratory ergonomics. We collected data from 380 participants and analyzed 11 anthropometric measurements, correlating them with 11 furniture dimensions. Two types of furniture were found and studied in different university computer laboratories: (1) a non-adjustable chair with a non-adjustable table and (2) an adjustable chair with a non-adjustable table. The mismatch calculation showed a significant difference between existing furniture dimensions and anthropometric measurements, indicating that 7 of the 11 existing furniture dimensions need improvement. The one-way ANOVA test with a significance level of 5\% also showed a significant difference between the anthropometric data and existing furniture dimensions. All 11 dimensions were determined to match students' anthropometric data. The proposed dimensions were found to be more compatible and showed reduced mismatch percentages for nine furniture dimensions and nearly zero mismatches for seat width, backrest height, and under the hood for both males and females compared to the existing furniture dimensions. The proposed dimensions of the furniture set with adjustable seat height showed slightly improved match results for seat height and seat-to-table clearance, which showed zero mismatches compared with the non-adjustable furniture set. The table width and table depth dimensions were suggested according to Barnes and Squires' ergonomic work envelope model, considering hand reach. The positions of the keyboard and mouse are also suggested according to the work envelope. The monitor position and viewing angle were proposed according to OSHA guidelines. This study suggests that the proposed dimensions can improve comfort levels, reducing the risk of musculoskeletal disorders among students. Further studies on the implementation and long-term effects of the proposed dimensions in real-world computer laboratory settings are recommended.
\end{abstract}


\end{frontmatter}

%\linenumbers

\section{Introduction}
\label{sec:introduction}
Ergonomics plays a crucial role in ensuring safety, health, and performance in various settings, including educational institutions, where efforts are made to optimize work and study environments for both teachers and students to enhance productivity and minimize the risk of musculoskeletal discomfort \cite{rudolf_evaluating_2009}. Integrating ergonomic principles into education can improve the quality and increase productivity. For instance, research has shown that sitting posture and positioning significantly influence typing and handwriting performance \cite{parush_ergonomic_1998}. Designing ergonomic solutions often relies on tools such as anthropometry, which involves measuring the size, shape, and capabilities of the human body for task performance \cite{bravo_literature_2018, mokdad_anthropometrics_2009}.

Several studies have proposed methods and guidelines for designing ergonomic classroom furniture tailored to students of different ages, providing relevant anthropometric dimensions and data for such designs \cite{arpaci_student_2016,fidelis_anthropometric_2022}. Among the various models for ergonomic furniture design, the 'design for adjustable range' model is widely recommended, as it considers a range of anthropometric measures of end-users, typically between the 5th and 95th percentiles of the user population \cite{fasulo_study_2019}. 

The motivation for this study stems from the growing recognition of the importance of ergonomic design for enhancing human productivity and well-being. Numerous studies have demonstrated the positive impact of ergonomically designed furniture on comfort, health, and performance, especially in settings where individuals spend prolonged periods of time seated at desks, such as offices and educational institutions. As computers and digital technology continue to permeate various aspects of academic and professional life, ensuring optimal ergonomic conditions in computer laboratories has become increasingly crucial. However, while ergonomic guidelines and standards exist for general office environments, tailored guidelines for computer laboratories, particularly for university students, are lacking. 

This study addresses this gap by proposing anthropometric-based furniture dimensions specifically designed for university students. This research aims to optimize furniture design in university computer labs to better accommodate students' physical characteristics.

The main goal of this study was to assess the current status of computer laboratories in Bangladesh by gathering anthropometric data from students. We sought to identify potential mismatches between anthropometric measurements and computer lab furniture dimensions to establish better design parameters for university students. By considering relevant anthropometric measurements of students sitting, our aim was to minimize potential mismatches and prolonged health issues. Additionally, this study aims to assess and compare the potential mismatch between two existing computer tables and chair sets. We proposed standardized design criteria and dimensions based on anthropometry and student safety guidelines.

Implementing anthropometric-based furniture dimensions tailored for university students can significantly enhance the ergonomic design of computer laboratories. By reducing potential mismatches between students' physical measurements and furniture dimensions, this research contributes to the creation of safer, more comfortable, and more productive learning environments. These optimized furniture dimensions have the potential to minimize the risk of musculoskeletal disorders (MSDs) and discomfort among students, thereby promoting their overall well-being and academic performance. Furthermore, by providing tailored guidelines for computer laboratory furniture design, this study offers practical insights that can inform future research and guide the development of ergonomic standards in educational institutions.

\section{Literature Review}
Hitka and Gejdoš analyzed a sample of Slovak university students from 2000-2006 and 2018-2023 for 25 anthropometric measures relevant to workspace design. They showed a decrease in Slovakia's adult population trend, with significant changes observed only in body weight among young adults \cite{hitka_m_decrease_2024}. Kang et al. investigated changes in body dimensions over time in Western (US) and Eastern (Korean) populations, focusing on the dimensions pertinent to automobile driver seat design. Data from the ANSUR and Size Korea datasets at two time points showed a consistent increase in body dimensions for both sexes in both populations \cite{kang_cross-cultural_2024}. Rababah and Etier showed that anthropometric measures in children are vital for designing child-centric furniture, tools, and toys. They analyzed a dataset of 354 children from Jordan across six age groups (six months to nine years), including 23 anthropometric measurements \cite{rababah_dataset_2024}. Goleij et al. (2024)  reviewed anthropometric changes worldwide over 30 years and analyzed 132 articles. They found significant increases in various body dimensions, emphasizing the need for regular updates and adaptations in work environments and equipment based on new anthropometric data \cite{goleij_investigating_2024}. Hitka et al. (2022) determined the ideal dimensions of wooden chairs for the adult bariatric subpopulation of Slovakia. By tracking long-term anthropometric changes, they were able to identify the body height of 95\% of adult males \cite{hitka_dimensional_2022}.

Laeser et al. recognized the necessity for furniture designed for 6th and 8th-grade children. In their study, students from these grades performed keyboarding and mousing tasks at two computer workstations. One workstation was a conventional desktop setup, and the other featured an adjustable keyboard. The study found that the overall posture scores of the students improved when using the adjustable workstation, as indicated by the Rapid Upper Limb Assessment \cite{laeser_effect_1998}. Thariq found that the current side-mounted desktop chair designs are not suited to fulfilling the postural and comfort requirements of university students in their learning environment. They conducted an anthropometric survey of university students in Sri Lanka. The survey aimed to create fixed-type and side-mounted desktop chairs and evaluate their comfort levels based on the design dimensions \cite{thariq_designing_2010}. Taifa and Desai recognized the issue of poorly designed furniture for engineering students in India. The authors employed a health survey, specifically an ergonomic assessment, to gather and analyze the data in their study. Based on their findings, the authors proposed the design of adjustable classroom furniture as a solution to reduce MSDs \cite{taifa_anthropometric_2017}. Hoque et al. found that classroom furniture does not consider ergonomics. The authors conducted an anthropometric survey of 500 Bangladeshi university students, followed by a mismatch analysis. Based on their findings, the authors recommended the use of ergonomically fit furniture~\cite{hoque_ergonomic_2014}. Joshi et al. addressed the significant rise in the design of computer workstations and the associated health and safety issues for users. They conducted postural assessment using the participative ergonomic technique of Rapid Upper Limb Assessment (RULA). The study revealed that most computer users lacked adequate information regarding the positioning of computer workstations, and a substantial number of computer operators reported issues with their upper extremities \cite{joshi_computer_2015}.

Kahya investigated the congruence between school furniture dimensions and students' anthropometric measurements. They assessed nine anthropometric dimensions of 225 students from nine departments. The results indicated a significant mismatch, with a seat height discrepancy of 44.45\% discrepancy, seat depth of 100\% discrepancy, and desk height of 21.28\% discrepancy. This study lays the groundwork for further exploration of classroom ergonomics in university settings \cite{kahya_mismatch_2019}. Abd Rahman et al. developed an anthropometric database tailored to Malaysian operators, focusing on sitting and standing dimensions. The database was constructed from the measurements of 146 male and 168 female participants aged 18-45 years. Thirty-six anthropometric dimensions were chosen to compare their availability across the four countries. This knowledge of population variations is crucial for designing workstations and facilities that accommodate the needs of the industrial environment \cite{abd_rahman_anthropometric_2018}. Langová et al. presented a method for designing chair dimensions based solely on the height range of intended users using seven equations describing the relationship between key design dimensions and human height or height range \cite{langova_mismatch_2021}. Sydor and Hitka linked chair dimensions to anthropometric measurements, simplifying access to data by establishing average body proportions for adults. Through seven equations, they related chair design dimensions to human height, thereby facilitating design considerations \cite{sydor_chair_2023}.

Parvez et al. identified that poorly fitting furniture could contribute to various MSDs and discomfort. To delve into this issue, they surveyed ten primary schools in Bangladesh, revealing a notable disparity between furniture dimensions and student body measurements. Moreover, this study proposes furniture dimensions that could notably decrease the mismatch percentage among students from 90\% to 10\% \cite{parvez_design_2018}.  Shohel Parvez et al. acknowledged that inadequately designed furniture could potentially exacerbate MSDs. They conducted a survey of 400 university students (250 males and 150 females) who volunteered for the study. The Standard Nordic Musculoskeletal Questionnaire was used to evaluate MSDs. The authors recommended that modifying the design of academic furniture could help alleviate or prevent MSD \cite{shohel_parvez_assessment_2022}. Moradi et al. conducted an ergonomic risk assessment of auto mechanics. This study found that back and waist circumference had the highest prevalence of work-related MSDs \cite{moradi_reba_2017}. Kibria and Rafiquzzaman developed a self-reported Nordic ergonomic assessment questionnaire to identify discrepancies between furniture dimensions and anthropometric measurements. The proposed ergonomically designed computer workstation, informed by anthropometric measurements and guidelines, can potentially alleviate MSDs among university teachers. This study could significantly influence ergonomic furniture design in universities and other organizations by addressing computer-related ergonomic challenges \cite{kibria_ergonomic_2019}.

As computer use grows, the risks of physical strains and eye issues increase. To combat this, various international standards like the International Organization for Standardization~\cite{ISO9241-5} and guidelines from organizations such as the Occupational Safety and Health Administration~\cite{osha2008} offer recommendations for ergonomic workstation design. These focus on factors such as the chair and desk setup, monitor positioning, and lighting to reduce strain and promote better health during computer use. The  International Organization for Standardization standard specifies basic dimensions for furniture used in office work, including tables and chairs, to promote user comfort and safety \cite{ISO5970-1979}. Additionally, The European Committee for Standardization (2020) provided guidelines for the ergonomic design of workstations, including layout and postural requirements for tasks involving visual display terminals (VDTs). This standard offers valuable insights into the workstation layout and postural requirements, which are crucial considerations in computer laboratory furniture design \cite{ISO9241-5}. Additionally, the European Committee for Standardization specifies the functional dimensions of the chairs and tables used in educational institutions. This European standard outlines ergonomic furniture requirements to promote students' health and well-being \cite{EN1729-1-2020}. European Committee for Standardization (2012) provides additional safety and ergonomic requirements for educational furniture, complementing \cite{EN1729-1, EN1729-2-2012}. Similarly, the European Committee for Standardization (2000) sets safety and ergonomic requirements for office chairs, ensuring user comfort and reducing the risk of musculoskeletal disorders \cite{EN1335-1-2000}. Incorporating the recommendations outlined in these standards ensures that computer laboratory furniture is ergonomically optimized for university students, minimizing the risk of MSDs and promoting student comfort and productivity.

Current literature on ergonomics and its impact on user health and performance has identified several research gaps that warrant further investigation. First, comprehensive research on interdisciplinary education in design and its impact on student learning outcomes is lacking. Second, few studies have been conducted on the impact of ergonomic furniture design on children's posture and health. Third, there is a shortage of research on university students' ergonomic needs, particularly in different countries and cultural contexts. Fourth, research on the effectiveness of ergonomic interventions in reducing the risk of MSDs and other ergonomic-related issues in educational settings is lacking. Fifth, limited research has been conducted on the impact of ergonomic interventions on work satisfaction, productivity, and overall well-being in various educational settings. Sixth, there is a shortage of research on the long-term effects of ergonomic interventions on students' health, performance, and retention. Finally, studies on the cost-effectiveness of ergonomic interventions and their potential return on investment in different occupational settings are lacking. Addressing these research gaps will contribute to the development of more effective ergonomic interventions and overall well-being in various occupational settings. However, there has been a limited emphasis on the design of furniture and computer labs for undergraduate education in universities, and there is a lack of established standards for computer lab furniture in Bangladeshi universities. Bangladeshi universities typically use desktops, chairs, and tables as computer laboratory furniture. The prolonged use of these improperly designed furniture pieces could exacerbate MSDs in young adults. Hence, ergonomically fitted furniture is crucial. Failure to address these issues promptly could result in undetected severe health problems among the students.

\section{Materials and Methods}
\label{sec:methodology}

\subsection{Problem Identification}
In our study, we sought to assess the ergonomic risks that students face due to prolonged sitting at their desks and the potential development of Work-related MSDs (WMSDs). We developed a comprehensive self-reported questionnaire based on the Nordic Musculoskeletal Questionnaire (NMQ), which has been extensively used and validated in previous studies. The questionnaire was designed to gather information on the frequency and severity of musculoskeletal symptoms experienced by teachers, focusing on areas such as the neck, shoulders, back, and wrists, which are commonly affected by prolonged sitting and poor ergonomics. Participants were asked to report the frequency of musculoskeletal symptoms they experienced, classifying them as ``constantly" (most of the time during the day), ``occasionally" (two to four times a month), or ``frequently" (more than four times a month). We also asked participants whether they had experienced musculoskeletal pain in the past 12 months. This allowed us to assess the prevalence of musculoskeletal pain among teachers and identify any data patterns or trends. Approximately 78.42\% of students reported experiencing MSD-related pain. In this group, 77.67\% of the patients were male, and 81.25\% were female. To better understand how these factors might influence the prevalence and severity of musculoskeletal symptoms, we collected demographic information, such as age, sex, and level of study. The questionnaire was administered to a sample of students of Khulna University of Engineering \& Technology (KUET). The collected data were analyzed to identify any significant associations between ergonomic factors, such as prolonged sitting and poor posture, and the development of WMSDs. The reliability test for the NMQ yielded a Cronbach's alpha value of 0.77, indicating an ``acceptable" level of internal consistency. This value falls within the 95\% confidence interval of 0.733 and 0.803, suggesting that if the survey were repeated with different samples, the actual value of Cronbach's alpha would likely fall within this range 95\% of the time.

\subsection{Sample Collection}
A subset of students was chosen from two departments at KUET, Khulna, Bangladesh. Equation \ref{eq1} was used to establish the necessary sample size.

\begin{equation}
    n = \frac{N}{1 + Ne^2}
\label{eq1}
\end{equation}

where $n$ represents the sample or population size, $N$ denotes the total population, and $e$ refers to the desired precision level set at 5\% with a 95\% confidence level.

According to KUET's official database, there are currently 5,240 students enrolled, comprising 4,318 males and 922 females. This yielded a male-to-female ratio of approximately 4.68:1. According to Equation \ref{eq1}, a minimum of 372 or more data points should be collected. This study gathered anthropometric data from 300 males and 80 females, representing up to 7,000 students. The students' ages ranged from 17 to 26 years. We collected data from June 2023 to September 2023.

All participants provided written consent for the study after being briefed on its aims. The Office of the Director for Research \& Extension, KUET, approved the experimental procedures conducted according to the principles stated in the Declaration of Helsinki, bearing the authorization code KUET/DRE/2023/15(6). Participation in the survey was voluntary, and participants had the right to withdraw without penalties. The confidentiality of participants' responses was strictly maintained, and all data collected were anonymized and stored securely. The dataset, available for further exploration, can be found in ``Anthropometric Data of KUET students" (\href{https://dx.doi.org/10.17632/kw7fd465v7.2}{https://dx.doi.org/10.17632/kw7fd465v7.2}) \cite{jahin_anthropometric_2024}.

\subsection{Anthropometric Measurements}
\begin{figure*}[!ht]
\centering
\includegraphics[width=0.6\linewidth]{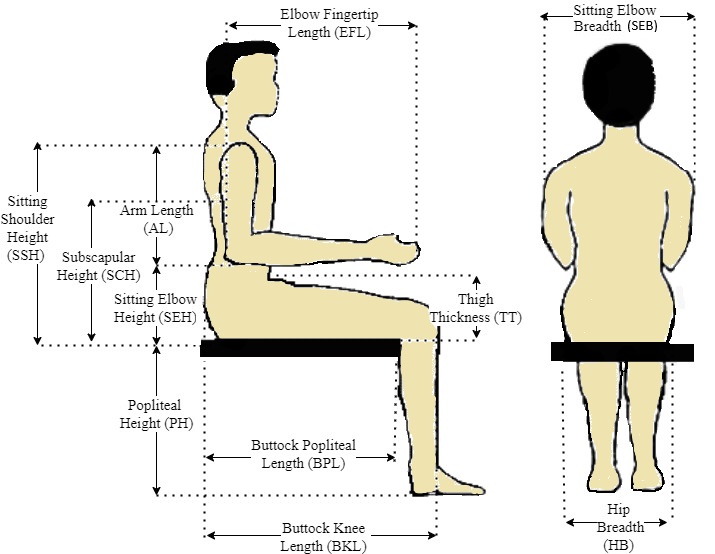}
\caption{Anthropometry of the participants, (a) right-side view (on the left), (b) backside view (on the right).}
\label{fig:3.2}
\end{figure*}

The 11 anthropometric measurements described in Figure \ref{fig:3.2} were obtained in this study. These dimensions were measured using definitions employed in earlier studies \cite{kroemer_engineering_1997}. A chair and Harpenden Anthropometer were used to determine the seating dimensions, and sitting assistance was used to support participants in maintaining a straight posture with their feet flat on the floor. Before data collection, rigorous calibration of all equipment was conducted in accordance with the established standards to maintain the accuracy and precision of the measurements. Throughout the trial, the participants were instructed to maintain a straight posture on a flat-surfaced seat with their knees positioned at a $90^\circ$ angle. Standardized posture is crucial for the consistency and comparability of anthropometric data. Participants were briefed about the study objectives, and precise postures were maintained during anthropometric measurements. All measurements were documented in millimeters (mm). Anthropometric measurements were performed by trained measurers following standardized techniques as outlined by the International Society for the Advancement of Kinanthropometry (ISAK) guidelines. All measurements were performed twice to ensure accuracy and reliability. The average of the two measurements was used for the data analysis.

\subsection{Furniture Dimensions}
\begin{figure*}[!ht]
\centering
\begin{subfigure}{.5\linewidth}
    \includegraphics[width=\linewidth]{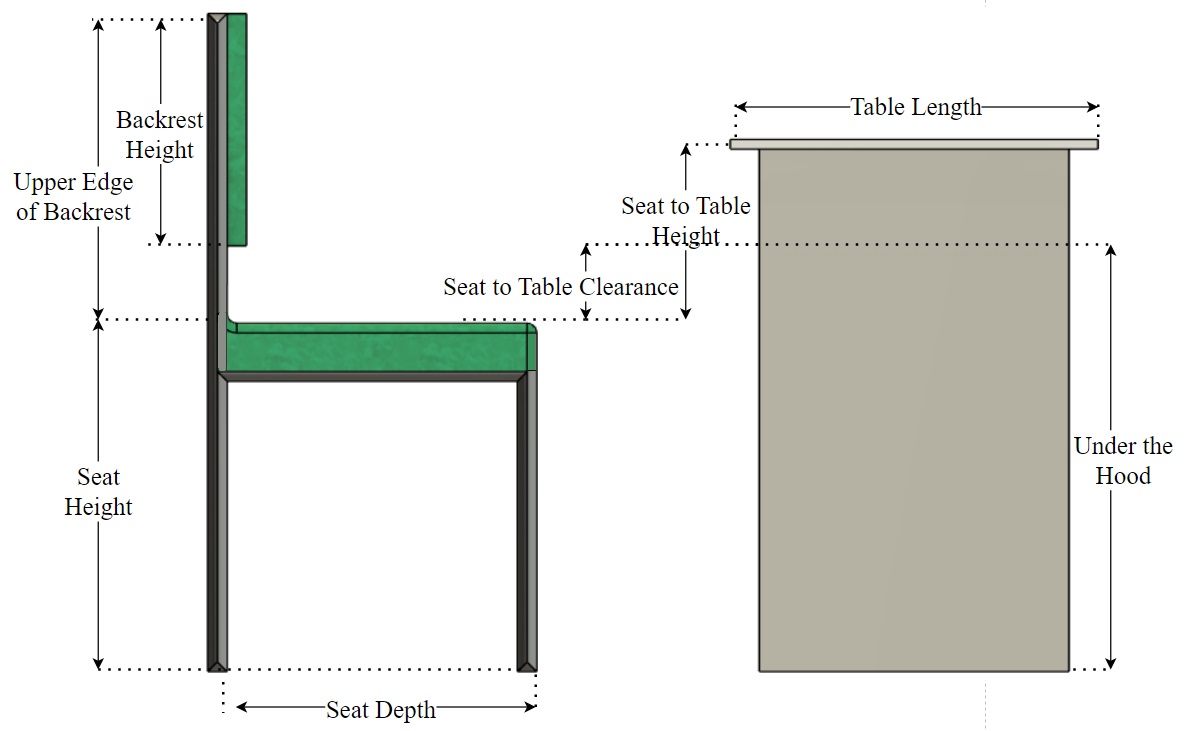}
    \caption{Left-side view}
\end{subfigure}
\begin{subfigure}{.45\linewidth}
    \includegraphics[width=\linewidth]{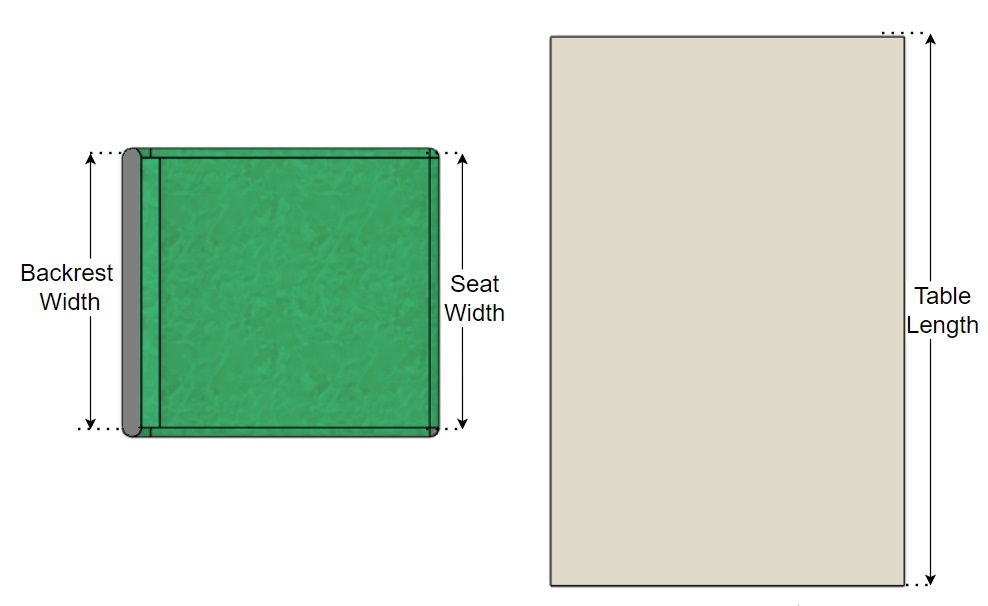}
    \caption{Top view}
\end{subfigure}
\caption{Furniture’s measurement: (a) Left-side view and (b) Top view.}
\label{fig:3.3}
\end{figure*}

\begin{figure*}[!ht]
\centering
\begin{subfigure}{0.5\linewidth}
    \includegraphics[width=1\linewidth]{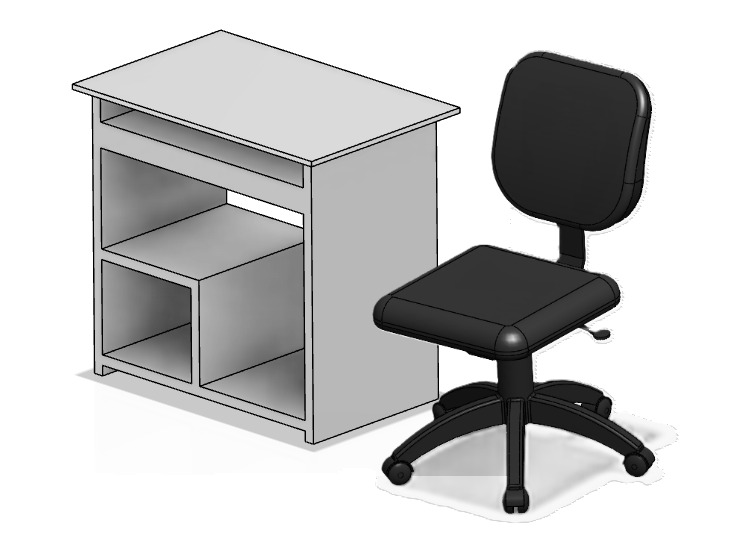}
    \caption{Table with adjustable chair}
    \label{fig:3.4a}
\end{subfigure}
\begin{subfigure}{0.45\linewidth}
    \centering
    \includegraphics[width=1\linewidth]{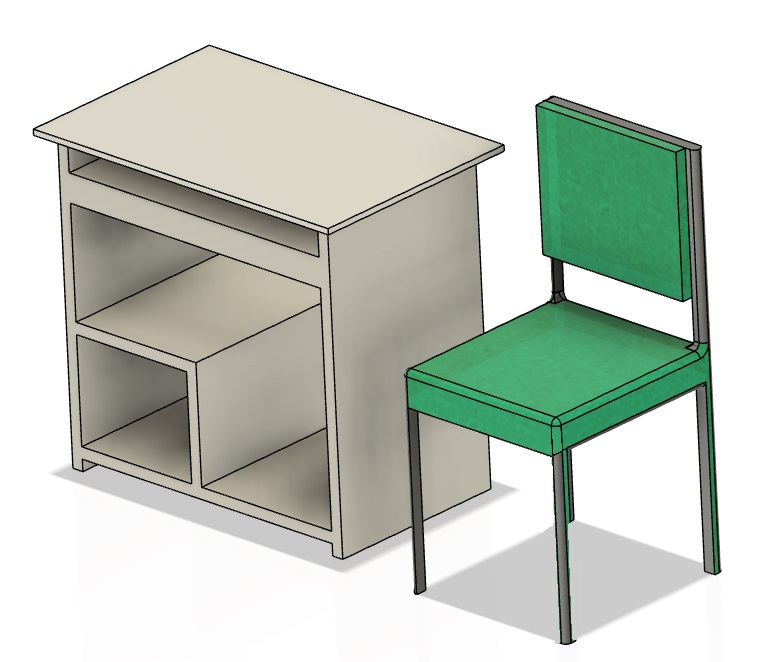}
    \caption{Table with non-adjustable chair}
    \label{fig:3.4b}
\end{subfigure}
\caption{Two types of furniture set, (a) Table with adjustable chair, (b) Table with non-adjustable chair.}
\label{fig:3.4}
\end{figure*}

The dimensions used to design the furniture are shown in Figure \ref{fig:3.3}. This study evaluated two types of university furniture sets used in KUET computer laboratories. One was a `table with adjustable chairs,' and the other was a `table with non-adjustable chairs,' both commonly used in computer labs at KUET. Figure \ref{fig:3.3} depicts the computer lab furniture design parameters, and Figure \ref{fig:3.4} shows the isometric view of two existing furniture sets. As such, this study measured 11 design parameters of existing computer furniture.

\subsection{Furniture Dimensions and Anthropometric Data Mismatch}
Considering the students' anthropometric measurements when designing and evaluating computer lab furniture is essential. This involves the use of relevant anthropometric estimations and ergonomic standards to determine the acceptable range for each furniture measurement.

Popliteal height (PH) and seat height (SH) are typically connected. The SH should provide knee flexion to align the lower legs with the body's craniocaudal axis at a $30^\circ$ angle \cite{jfm_revision_2003}. According to \cite{parcells_mismatch_1999}, PH should be higher than SH. This means that the shin-thigh position should be between $95^\circ$ and $120^\circ$, and the lower leg should create an angle with a vertical axis of $5^\circ$ to $30^\circ$. While an unduly low SH can strain the ischial tuberosities, an extremely high SH can increase the pressure on the bottom of the knee and nerves, thereby decreasing blood circulation \cite{hoque_ergonomic_2014}. The SH should be less than 40 mm, or 88\% of the PH, to prevent excessive pressure on the buttocks \cite{jfm_revision_2003}. The PH measurement considered shoe height by adding 30 mm. Therefore, Equation \ref{eq2} can be used to represent the relationship between PH and SH:

\begin{equation}
    (PH + 30)\cos(30^\circ) \leq SH \leq (PH + 30)\cos(5^\circ)
    \label{eq2}
\end{equation}

To allow for both furniture flexibility and essential motions, the seat width (SW) should be broad enough to accommodate people with the widest hip breadth (HB)~\cite{evans_design_1988,sanders_human_1993}. A proper fit is suggested by previous research, and the 95th percentile of HB is a good place to start when measuring SW \cite{evans_design_1988,occhipinti_criteria_1993,oborne_ergonomics_1995,sanders_human_1993}. The SW should be wider than the HB. Previous research has suggested that the optimal SW should be within the range of 10\% to 30\% of HB \cite{dianat_classroom_2013,gouvali_match_2006}. Consequently, the link between the HB and SW can be established using Equation \ref{eq3}: 

\begin{equation}
    (1.10 \times HB) \leq SW \leq (1.30 \times HB)
    \label{eq3}
\end{equation}

There is a strong relationship between seat depth (SD) and buttock popliteal length (BPL), with the BPL frequently serving as a guide for SD development and evaluation. The fifth percentile of the BPL should be used as the basis for the seat depth, according to several studies \cite{sanders_human_1993,oborne_ergonomics_1995,panagiotopoulou_classroom_2004}. This guarantees that the seat backrest can support the lumbar spine without compressing the popliteal surface. The seat may not support the thigh when the SD is much shorter than the BPL \cite{castellucci_mismatch_2010}. However, the user might be unable to use the backrest to support their lumbar spine if the SD is larger than the BPL \cite{milanese_school_2004}. Accordingly, when the SD is either 95\% of BPL or less, there is a mismatch between the SD and BPL \cite{afzan_mismatch_2012,altaboli_anthropometric_2015,dianat_classroom_2013}. Equation \ref{eq4} can be used to express the connection between BPL and SD:

\begin{equation}
    (0.80 \times BPL) \leq SD \leq (0.95 \times BPL)
    \label{eq4}
\end{equation}

According to several studies, sitting elbow height (SEH) is a critical factor in Seat to Table Height (STH) determination \cite{bendix_how_1986,milanese_school_2004,jfm_revision_2003}. Developing an appropriate Table Height (TH) requires careful consideration of shoulder biomechanics, including shoulder flexion angles between $0^\circ$ and $25\circ$ and abduction angles between $0^\circ$ and $20^\circ$ \cite{agha_school_2010,altaboli_anthropometric_2015,gouvali_match_2006}. Furthermore, keeping the arms on the desk significantly reduces strain on the spine. Consequently, according to ref \cite{poulakakis_model_1998}[50 Pheasant, 51 Poulakakis], the STH should be 3-5 cm higher than the SEH. A synchronized measure was created to verify that the SEH is the most important TH parameter and that the TH height should not increase above 50 mm above the SEH. Equation \ref{eq5} expresses the link between TH and SEH:

\begin{equation}
    SEH \leq TH \leq (SEH+50)
    \label{eq5}
\end{equation}

The flexibility of the arms and upper trunk depends on the appropriate backrest height (BH). It is recommended that the BH value should not be higher than that of the scapula \cite{evans_design_1988}. Therefore, an ideal BH is between 60\% and 80\% of sitting shoulder height (SSH) \cite{agha_school_2010,dianat_classroom_2013,castellucci_equations_2015}. BH dimensions should also be less than or equal to the top margin of the scapula. This creates a match criterion, and Equation \ref{eq6} shows how BH and SSH relate to each other:

\begin{equation}
    (0.60 \times SSH) \leq BH \leq (0.80 \times SSH)
    \label{eq6}
\end{equation}

The seat-to-table clearance (STC) should be sufficiently large to provide adequate space for leg movement and for the user to move the chair below the table \cite{garcia-acosta_definition_2007,jfm_revision_2003}. If the STC is greater than the thigh thickness (TT), it is considered a standard dimension \cite{parcells_mismatch_1999}. However, the ideal dimension of STC should be 20 mm higher than the knee height \cite{gouvali_match_2006}. Therefore, a match criterion was developed, and the relationship between SDC and TT is shown in Equation \ref{eq7}:

\begin{equation}
    (TT+20) < STC
    \label{eq7}
\end{equation}

The user should have support for simple access to the seat, leg mobility, and getting off the seat from under the hood (UTH). In addition, the design must incorporate SH, TT, and an additional 20 mm of space to allow for a shift in leg posture and the lowest standard dimension of UTH \cite{perez-gosende_anthropometry-based_2017}. However, according to \cite{gouvali_match_2006}, the highest dimension of UTH cannot be greater than the difference between the maximum dimension of TH and the table thickness, which is 30mm. Consequently, the link between UTH, PH, and TT yields a match equation, as shown in Equation \ref{eq8}:

\begin{multline}
    (SH + TT + 30) \leq UTH \leq (SEH + [(PH +30)\cos(5^\circ)]\\ + 0.1483AL - 30)
    \label{eq8}
\end{multline}

Hip breadth (HB) is a key metric in the backrest width (BW) design process \cite{thariq_designing_2010}. Taifa and Desai recently suggested that HB is a pertinent anthropometric measurement for BW design~\cite{taifa_anthropometric_2017}. Consequently, an equation was created, and Equation \ref{eq9} illustrates the relationship between HB and BW: 

\begin{equation}
    BW \geq HB
    \label{eq9}
\end{equation}

The table length (TL) is crucial for providing sufficient room for shifting lower body postures or movements. Previous research has suggested that TL should be higher than buttock knee length (BKL) \cite{perez-gosende_anthropometry-based_2017}. Thus, Equation \ref{eq10} defines the TL minimal limit.

\begin{equation}
    TL \geq BKL
    \label{eq10}
\end{equation}

The table depth (TD) and sitting elbow breath (SEB) anthropometric measurements were connected. The partial widths of the arm length (AL) and SEB should be included in the lower bound of the table depth (TD) \cite{perez-gosende_anthropometry-based_2017}. Sufficient room is needed for a table so that one can enter and leave the seat. TD thus permits elbow abduction at $20^\circ$ with a 20 mm margin \cite{afzan_mismatch_2012,agha_school_2010,altaboli_anthropometric_2015,gouvali_match_2006}. Equation \ref{eq11} considers these factors as follows:

\begin{equation}
    (0.5SEB+0.342AL+20) \leq TD \leq EFL
    \label{eq11}
\end{equation}

Another important anthropometric criterion for evaluating the upper edge of the backrest (UEB) is the sub-scapular height (SCH) \cite{bendix_how_1986}. Numerous studies have indicated that the scapula and arm are immobile if the SCH is smaller than the UEB \cite{castellucci_mismatch_2010,garcia-acosta_definition_2007,gouvali_match_2006}. Equation \ref{eq12}, therefore, depicts the link between SCH and UEB.

\begin{equation}
    UEB \leq SCH
    \label{eq12}
\end{equation}

\subsection{Level of Compatibility}
Some equations have single-directional limitations, while others have dual constraints to assess how well participants' anthropometric measurements match the furniture dimensions. Match and mismatch categories were designed to evaluate compatibility in single-directional connections. Three categories were established for dual-directional relationships: (a) match (where the anthropometric measurement falls between the limits), (b) low mismatch (where the anthropometric measurement falls below the maximum limit of the relationship), and (c) high mismatch (where the minimum limit of the relationship exceeds the anthropometric measurement).

\subsection{Software and Statistical Tools}
IBM SPSS Statistics 26 and Microsoft Excel 2021 were used for statistical analysis. For the graphical 3D prototype design, we used the Autodesk Fusion 360. For visualizations related to anthropometric data distribution and correlation analysis, Python 3.10.

\section{Results}
\label{sec: Results}
\subsection{Analysis of Anthropometric Measurements} 
Eleven anthropometric measures were used to compute the descriptive statistics. The following values were estimated to create a standard database (Table \ref{tab:4.1}): minimum (min), maximum (max), mean, standard deviation, and various percentile values (5th, 50th, and 95th percentiles). Anthropometric measurements of the male and female participants in this study were computed independently.

\subsection{Assessment of Intra- and Inter-Class Reliability}
Intra-class reliability was assessed by comparing the measurements taken by the same measurer on two separate occasions, whereas inter-class reliability was assessed by comparing measurements taken by different measures. Intra- and inter-class reliability for the anthropometric measurements was assessed using a two-way mixed effects model, where people's effects were considered random, and the measure effects were fixed. The assessment of intra- and inter-class reliability is crucial for ensuring the precision and consistency of our data. 

In this study, we obtained high intra-class correlation coefficients ($ICC_1$) for both single measures ($ICC_1$ = 0.973, 95\% CI: 0.966 - 0.979) and average measures ($ICC_1$ = 0.986, 95\% CI: 0.983 - 0.989). These findings indicate a high level of agreement and consistency between the repeated measurements. The inter-class reliability ($ICC_2$) was assessed by comparing the measurements taken by different researchers. Similarly, we found high $ICC_2$ values for both single measures ($ICC_2$ = 0.944, 95\% CI: 0.931-0.956) and average measures ($ICC_2$ = 0.971, 95\% CI: 0.964 - 0.977). These results indicate a high level of agreement and consistency between measurements taken by different researchers.

\subsection{Analysis of Existing Furniture Dimensions}
Table \ref{tab:4.2} lists the 11 existing furniture dimensions of the measured computer furniture. Type-1 indicates a table with a non-adjustable chair, whereas type-2 indicates a table with an adjustable chair.

\begin{table*}[!ht]
\centering
\caption{Existing two types of furniture dimensions}
\label{tab:4.2}
\begin{tabular}{| l | l | l |}
\hline
\textbf{Dimensions } & \textbf{Type-1 Furniture set (mm)} & \textbf{Type-2 Furniture set (mm)} \\
\hline
SH  & 457 & 432 - 533 \\
\hline
SW & 394 & 457 \\
\hline
SD & 406 & 432 \\
\hline
BH  & 305 & 305 \\
\hline
BW  & 356 & 394 \\
\hline
UEB  & 406 & 406 \\
\hline
STH  & 241 & 229 - 330 \\
\hline
STC & 89 & 95 - 197 \\
\hline
UTH  & 546 & 629 \\
\hline
TL & 483 & 457 \\
\hline
TD  & 749 & 749 \\
\hline
\end{tabular}
\end{table*}

\subsection{Mismatch Analysis}
To evaluate the degree of mismatch between the current library furniture and body measurements, equations \ref{eq2} through \ref{eq12} were applied. To find a mismatch, the writers examined into 11 aspects of computer furniture. Individual matches or mismatches were identified by comparing the dimensions of the furnishings with the body measurements of each participant. Finally, a percentage representing the measured match or mismatch for the sample population was presented.

%All the parameters are compared in Figure \ref{fig:4.4} to understand the scenario better. 

\begin{table*}[!ht]
\centering
\caption{Summary of the level of mismatch and match percentages of existing furniture dimensions of 1st type of furniture (non-adjustable chair and non-adjustable table)}
\label{tab:4.3}
\resizebox{\linewidth}{!}{%
\begin{tabular}{|l|l|l|l|l|l|} 
\hline
\textbf{Dimensions and anthropometry} & \textbf{Gender} & \textbf{Match (\%)} & \textbf{Lower mismatch\textbf{~(\%)}} & \textbf{Upper mismatch\textbf{~(\%)}} & \textbf{Total mismatch\textbf{~(\%)}} \\ 
\hline
\multirow{2}{*}{SH against PH} & Male & 86.67 & 13.33 & 0 & 13.33 \\ 
\cline{2-6}
 & Female & 12.5 & 87.5 & 0 & 87.5 \\ 
\hline
\multirow{2}{*}{SW against HB} & Male & 78.67 & 0 & 21.33 & 21.33 \\ 
\cline{2-6}
 & Female & 18.75 & 0 & 81.25 & 81.25 \\ 
\hline
\multirow{2}{*}{SD against BPL} & Male & 92 & 8 & 0 & 8 \\ 
\cline{2-6}
 & Female & 86.25 & 13.75 & 0 & 13.75 \\ 
\hline
\multirow{2}{*}{BH against SSH} & Male & 39 & 0 & 61 & 61 \\ 
\cline{2-6}
 & Female & 93.75 & 0 & 6.25 & 6.25 \\ 
\hline
\multirow{2}{*}{BW against HB} & Male & 70.33 & - & - & 29.67 \\ 
\cline{2-6}
 & Female & 10 & - & - & 90 \\ 
\hline
\multirow{2}{*}{UEB against SCH} & Male & 100 & - & - & 0 \\ 
\cline{2-6}
 & Female & 100 & - & - & 0 \\ 
\hline
\multirow{2}{*}{STH against SEH} & Male & 66.33 & 0.33 & 33.33 & 33.67 \\ 
\cline{2-6}
 & Female & 72.5 & 1.25 & 26.25 & 27.5 \\ 
\hline
\multirow{2}{*}{STC against TT} & Male & 0 & - & ~- & 100 \\ 
\cline{2-6}
 & Female & 0 & - & ~- & 100 \\ 
\hline
\multirow{2}{*}{UTH against TT, SEH, PH and AL} & Male & 0 & 0 & 100 & 100 \\ 
\cline{2-6}
 & Female & 0 & 0 & 100 & 100 \\ 
\hline
\multirow{2}{*}{TL against BKL} & Male & 0 & - & - & 100 \\ 
\cline{2-6}
 & Female & 2.5 & - & - & 97.5 \\ 
\hline
\multirow{2}{*}{TD against SEB, AL and EFL} & Male & 0 & 100 & 0 & 100 \\ 
\cline{2-6}
 & Female & 0 & 100 & 0 & 100 \\
\hline
\end{tabular}
}
\end{table*}

% Figure \ref{fig:4.5} compares all the parameters more explicitly.

\begin{table*}[!ht]
\centering
\caption{Summary of the level of mismatch and match percentages of existing furniture dimensions of 2nd type of furniture (adjustable chair and non-adjustable table)}
\label{tab:4.4}
\resizebox{\linewidth}{!}{%
\begin{tabular}{|l|l|l|l|l|l|} 
\hline
\textbf{Dimensions and anthropometry} & \textbf{Gender} & \textbf{Match (\%)} & \textbf{Lower mismatch (\%)} & \textbf{Upper mismatch (\%)} & \textbf{Total mismatch (\%)} \\ 
\hline
\multirow{2}{*}{SH against PH} & Male & 100 & 0 & 0 & 0 \\ 
\cline{2-6}
 & Female & 82.5 & 17.5 & 0 & 17.5 \\ 
\hline
\multirow{2}{*}{SW against HB} & Male & 43 & 57 & 0 & 57 \\ 
\cline{2-6}
 & Female & 96.25 & 3.75 & 0 & 3.75 \\ 
\hline
\multirow{2}{*}{SD against BPL} & Male & 49.67 & 50.33 & 0 & 50.33 \\ 
\cline{2-6}
 & Female & 38.75 & 61.25 & 0 & 61.25 \\ 
\hline
\multirow{2}{*}{BH against SSH} & Male & 39 & 0 & 61 & 61 \\ 
\cline{2-6}
 & Female & 93.75 & 0 & 6.25 & 6.25 \\ 
\hline
\multirow{2}{*}{BW against HB} & Male & 100 & - & - & 0 \\ 
\cline{2-6}
 & Female & 100 & - & - & 0 \\ 
\hline
\multirow{2}{*}{UEB against SCH} & Male & 100 & - & - & 0 \\ 
\cline{2-6}
 & Female & 100 & - & - & 0 \\ 
\hline
\multirow{2}{*}{STH against SEH} & Male & 100 & 0 & 0 & 0 \\ 
\cline{2-6}
 & Female & 100 & 0 & 0 & 0 \\ 
\hline
\multirow{2}{*}{STC against TT} & Male & 99.33 & - & - & 0.67 \\ 
\cline{2-6}
 & Female & 100 & - & - & 0 \\ 
\hline
\multirow{2}{*}{UTH against TT, SEH, PH and AL} & Male & 92 & 0 & 8 & 8 \\ 
\cline{2-6}
 & Female & 97.5 & 0 & 2.5 & 2.5 \\ 
\hline
\multirow{2}{*}{TL against BKL} & Male & 0 & - & - & 100 \\ 
\cline{2-6}
 & Female & 0 & - & - & 100 \\ 
\hline
\multirow{2}{*}{TD against SEB, AL and EFL} & Male & 0 & 100 & 0 & 100 \\ 
\cline{2-6}
 & Female & 0 & 100 & 0 & 100 \\
\hline
\end{tabular}
}
\end{table*}

\subsection{Statistical Test}
One-way ANOVA was used to determine whether there were any statistically significant differences between the means of three or more independent groups. In this study, the furniture dimensions were tested against anthropometric dimensions to determine any significant differences. The level of significance ($\alpha$) was set at 5\%, meaning that the result was considered statistically significant if the p-value was $\leq5\%$. The result was not statistically significant if the p-value was $>5\%$. The F-value measures the variance between groups compared with the variance within groups. The p-value is the probability of observing a result as extreme as obtained, assuming that the null hypothesis is true. If the p-value was less than the $\alpha$ value (5\% in this case), the null hypothesis was rejected, and it was concluded that there was a statistically significant difference between the groups. If the p-value was greater than the $\alpha$ value, the null hypothesis was not rejected, and it was concluded that there was no statistically significant difference between the groups.

The results showed that SH, SW, BH, UEB, STH, STC, and TL of furniture dimensions were rejected for the type-1 furniture set, meaning there were statistically significant differences between the groups. In contrast, SD and STH were accepted, indicating that there were no statistically significant differences between the groups (Table \ref{tab:4.5}). For the adjustable chair non-adjustable table, we considered both the highest and lowest limits of SH, STH, and STC. If the highest or lowest limit is accepted for males and females, we consider the parameter accepted.  Only the SH, SD, and STH dimensions were accepted according to the ANOVA test for adjustable chairs with non-adjustable tables. The remaining dimensions, SW, BH, BW, UEB, STC, and TL, were rejected (Table \ref{tab:4.6}).

\begin{itemize}
    \item \textit{$H_{0}$ (null hypothesis):} There is no effect or difference between anthropometric measurements and existing furniture dimensions.
    \item \textit{$H_{1}$ (alternative hypothesis):} There is an effect or difference between anthropometric measurements and existing furniture dimensions.
\end{itemize}

\subsection{Proposed Dimensions of Furniture}
\begin{figure*}[!ht]
\centering
\begin{subfigure}{0.45\linewidth}
    \includegraphics[width=\linewidth]{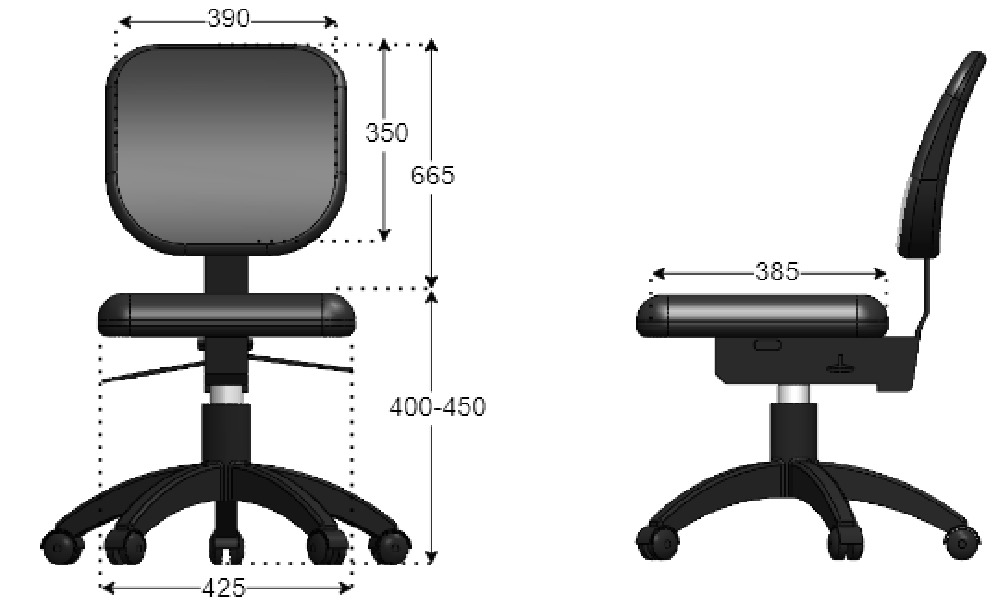}
    \caption{Adjustable chair}
    \label{fig:4.7a}
\end{subfigure}
\begin{subfigure}{0.45\linewidth}
    \includegraphics[width=\linewidth]{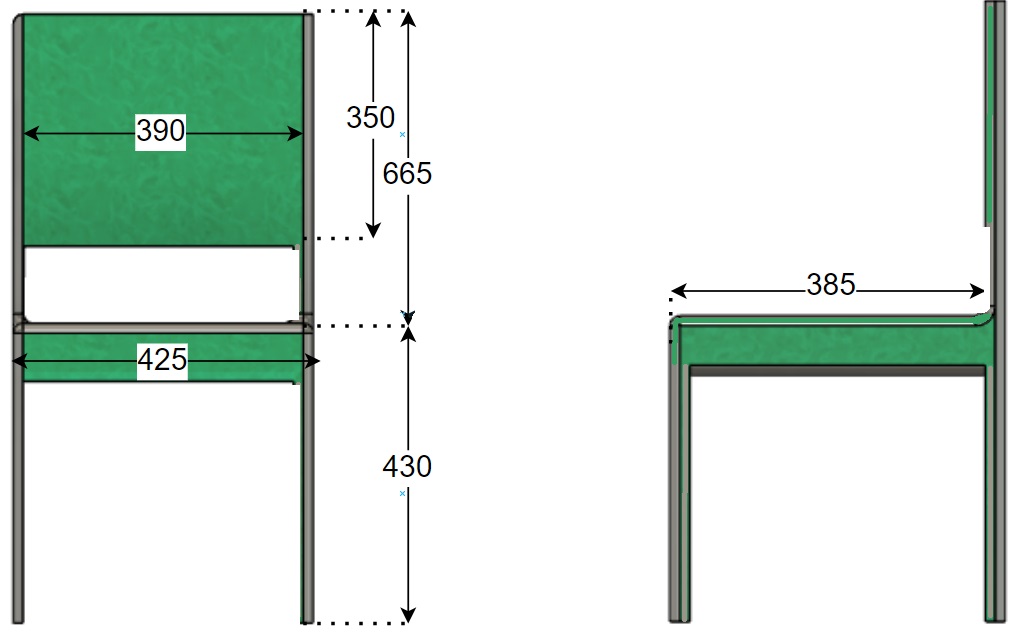}
    \caption{Non-adjustable chair}
    \label{fig:4.7b}
\end{subfigure}
\caption{Proposed dimensions for both types of chairs (in mm): (a) adjustable chair, (b) non-adjustable chair.}
\end{figure*}

Table \ref{tab:4.7} presents the proposed dimensions for the two types of furniture sets designed for university computer laboratories, along with a comparison with relevant International Standards (ISO/EN) for ergonomic furniture design. The dimensions include SH, SW, SD, BH, BW, UEB, STH, STC, and UTH. The type-1 and type-2 furniture sets have slightly different proposed dimensions, particularly for SH, STH, and STC. The table also indicates each dimension's corresponding International Standards (ISO/EN), providing a reference for optimal ergonomic design in university computer laboratories. The SH is suggested to be between 400 mm and 450 mm for adjustable chairs( Figure \ref{fig:4.7a}). The recommended SH is 430 mm for non-adjustable chairs, as shown in Figure \ref{fig:4.7b}. The proposed SW is 425 mm, SD is 385 mm, BH is suggested to be 350 mm, BW is 390 mm, UEB is proposed to be 465 mm, and UTH is 645 mm.

\begin{table*}[!ht]
\centering
\caption{Proposed dimensions of furniture sets and comparison with International Standards}
\label{tab:4.7}
\begin{tabular}{|l|>{\raggedright\arraybackslash}p{0.18\linewidth}|>{\raggedright\arraybackslash}p{0.18\linewidth}|>{\raggedright\arraybackslash}p{0.35\linewidth}|}
\hline
\textbf{Dimensions} & \textbf{Type-1 Proposed Dimensions (mm)} & \textbf{Type-2 Proposed Dimensions (mm)} & \textbf{Proposed dimensions according to International Standards (mm)} \\
\hline
SH & 430 & 400 - 450 & 410 - 490 (ISO 5970:1979) \\
\hline
SW & 425 & 425 &  440 - 480 (EN 1729-1:2020)\\
\hline
SD & 385 & 385 &  390 - 450 (EN 1729-2:2012)\\
\hline
BH & 350 & 350 &  340 - 400 (EN 1729-2:2012)\\
\hline
BW & 390 & 390 &  370 - 430 (ISO 5970:1979)\\
\hline
UEB & 465 & 465 &  450 - 530 (EN 1335-1:2000)\\
\hline
STH & 260 & 235 - 310 & 260 - 340 (EN 1729-1:2016-02)\\
\hline
STC & 200 & 95.25 - 200 & 95 - 225 (EN 1729-1:2016-02)\\
\hline
UTH & 645 & 645 & 625 - 705 (ISO 5970:1979)\\
\hline
\end{tabular}
\end{table*}

For non-adjustable chairs and tables, Table \ref{tab:4.8} presents the mismatch percentages of the proposed furniture measurements. The SH against PH showed 5\% and 12.5\% for males and females, respectively. The SW against HB showed zero mismatch for males and only 1.25 lower mismatch for females. The SD against BPL showed 7.33\% high mismatch and 0.33 \% low mismatch for males, whereas 0\% high mismatch and 1.25\% low mismatch for females. It is necessary to reduce the high mismatch for females as much as possible when designing a chair's SD. BH, BW, UEB, and STC showed a 100\% match for both males and females.

Compared with the previous mismatch percentages in university furniture, the proposed dimensions exhibited reduced mismatches for both males and females. The suggested measurements were more compatible than the previous dimensions, leading to improved comfort levels and reduced risks of MSDs for users.

\begin{table*}[!ht]
\centering
\caption{Proposed furniture mismatch analysis for type 1 (non-adjustable chair and table)}
\label{tab:4.8}
\resizebox{\linewidth}{!}{%
\begin{tabular}{|l|l|l|l|l|l|} 
\hline
\textbf{Dimensions and anthropometry} & \textbf{Gender} & \textbf{Match (\%)} & \textbf{Low mismatch (\%)} & \textbf{High mismatch (\%)} & \textbf{Total mismatch (\%)} \\ 
\hline
\multirow{2}{*}{SH against PH} & Male & 95 & 0 & 5 & 5 \\ 
\cline{2-6}
 & Female & 87.5 & 12.5 & 0 & 12.5 \\ 
\hline
\multirow{2}{*}{SW against HB} & Male & 100 & 0 & 0 & 0 \\ 
\cline{2-6}
 & Female & 98.75 & 1.25 & 0 & 1.25 \\ 
\hline
\multirow{2}{*}{SD against BPL} & Male & 92.33 & 0.33 & 7.33 & 7.67 \\ 
\cline{2-6}
 & Female & 98.75 & 1.25 & 0 & 1.25 \\ 
\hline
\multirow{2}{*}{BH against SSH} & Male & 100 & 0 & 0 & 0 \\ 
\cline{2-6}
 & Female & 100 & 0 & 0 & 0 \\ 
\hline
\multirow{2}{*}{BW against HB} & Male & 100 & - & - & 0 \\ 
\cline{2-6}
 & Female & 100 & - & - & 0 \\ 
\hline
\multirow{2}{*}{UEB against SCH} & Male & 100 & - & - & 0 \\ 
\cline{2-6}
 & Female & 100 & - & - & 0 \\ 
\hline
\multirow{2}{*}{STH against SEH} & Male & 87.5 & 8.75 & 3.75 & 12.5 \\ 
\cline{2-6}
 & Female & 87.5 & 8.75 & 3.75 & 12.5 \\ 
\hline
\multirow{2}{*}{STC against TT} & Male & 100 & - & - & 0 \\ 
\cline{2-6}
 & Female & 100 & - & - & 0 \\ 
\hline
\multirow{2}{*}{UTH against TT, SEH, PH and AL} & Male & 100 & 0 & 0 & 0 \\ 
\cline{2-6}
 & Female & 98.75 & 0 & 1.25 & 1.25 \\
\hline
\end{tabular}
}
\end{table*}

Table \ref{tab:4.9} presents the mismatch percentages of the proposed furniture measurements for the adjustable chairs and tables. Compared to the previous mismatch percentages in university furniture, the proposed dimensions exhibited reduced mismatches for males and females. The suggested measurements were more compatible than the previous dimensions, leading to improved comfort levels and reduced risks of MSDs for users.

\begin{table*}[!ht]
\centering
\caption{Proposed furniture mismatch analysis for type-2 (adjustable chairs with non-adjustable tables) furniture set}
\label{tab:4.9}
\resizebox{\linewidth}{!}{%
\begin{tabular}{|l|l|l|l|l|l|} 
\hline
\textbf{Dimensions and anthropometry} & \textbf{Gender} & \textbf{Match (\%)} & \textbf{Low mismatch (\%)} & \textbf{High mismatch (\%)} & \textbf{Total mismatch (\%)} \\ 
\hline
\multirow{2}{*}{SH against PH} & Male & 100 & 0 & 0 & 0 \\ 
\cline{2-6}
 & Female & 100 & 0 & 0 & 0 \\ 
\hline
\multirow{2}{*}{SW against HB} & Male & 100 & 0 & 0 & 0 \\ 
\cline{2-6}
 & Female & 98.75 & 0 & 1.25 & 1.25 \\ 
\hline
\multirow{2}{*}{SD against BPL} & Male & 92.33 & 0.33 & 7.33 & 7.67 \\ 
\cline{2-6}
 & Female & 98.75 & 1.25 & 0 & 1.25 \\ 
\hline
\multirow{2}{*}{BH against SSH} & Male & 100 & 0 & 0 & 0 \\ 
\cline{2-6}
 & Female & 100 & 0 & 0 & 0 \\ 
\hline
\multirow{2}{*}{BW against HB} & Male & 100 & - & - & 0 \\ 
\cline{2-6}
 & Female & 100 & - & - & 0 \\ 
\hline
\multirow{2}{*}{UEB against SCH} & Male & 100 & - & - & 0 \\ 
\cline{2-6}
 & Female & 100 & - & - & 0 \\ 
\hline
\multirow{2}{*}{STH against SEH} & Male & 100 & 0 & 0 & 0 \\ 
\cline{2-6}
 & Female & 100 & 0 & 0 & 0 \\ 
\hline
\multirow{2}{*}{STC against TT} & Male & 100 & -~ & -~ & 0 \\ 
\cline{2-6}
 & Female & 100 & -~ & -~ & 0 \\ 
\hline
\multirow{2}{*}{UTH against TT, SEH, PH and AL} & Male & 100 & 0 & 0 & 0 \\ 
\cline{2-6}
 & Female & 98.75 & 0 & 1.25 & 1.25 \\
\hline
\end{tabular}
}
\end{table*}

\section{Discussion}

The normal bell-shaped curves observed in Figures \ref{fig:4.1} and \ref{fig:4.2} indicate the reliability and consistency of the collected 11 anthropometric data collected. In statistics, a normal distribution is often considered a hallmark of reliable data, as it suggests that the measurements are evenly distributed around the mean, with fewer extreme values and more values clustered around the center. The presence of these curves in the data for both sexes indicated that the measurements were accurate and consistent without significant outliers or inconsistencies. This reliability is crucial in ergonomic research, as it ensures that the collected data accurately represent the population and can be used to make informed decisions about furniture design and ergonomics. Additionally, the normal distributions in both male and female datasets suggest that the collected measurements represent the general population rather than being skewed or biased toward one gender. This further enhances the reliability and validity of the dataset and supports its applicability to various settings and industries.

The high intra- and inter-class reliabilities of our study indicate that our anthropometric measurements are highly reliable and consistent. These findings enhance the validity and reliability of our data and provide confidence in the accuracy of our measurements.

In Table \ref{tab:4.3}, for the first type of furniture, the mismatch level is notably higher for male body measurements than for females in parameters, such as SH was 74.17\%, SW was 59.92\%, BW was 60.33\%, and SD was 5.75\% higher whereas BH was 54.75\%, and STH was 6.17\% lower. Only one parameter, UEB, was found to be a 100\% match for both sexes. The results were particularly poor for table-related parameters, indicating that furniture was poorly fitted for users in these dimensions. The results for the second type of furniture in Table \ref{tab:4.4} show significantly lower mismatch levels than those for the first type of furniture for SH, SW, BW, STC, and UTH. BW, UEB, and STH showed zero mismatches for this furniture set, and no improvements were required. However, the mismatch percentage is still sufficiently large enough for the remaining eight parameters, which still require improvements.  However, the outcomes for the table-related parameters remain subpar, indicating that furniture is still not well-suited for users in these dimensions. 

After the statistical test, the decisions were more understandable, and 11 furniture dimensions were proposed based on 11 anthropometric data collected. The proposed dimensions for UTH showed a 100\% match for both furniture sets. The SH was proposed based on the 5th percentile of the female popliteal height with 30 mm shoe clearance. The proposed SW is 425 mm, based on the 95th percentile of HB in females. The recommended SD is 385 mm, as determined by the 5th percentile of female BPL. BH is suggested to be 350 mm, considering the 5th percentile of female SSH. The recommended BW is 390 mm, based on the 95th percentile of male SEB. The UEB is proposed to be 465 mm, considering the minimum subscapular height of the female participants. The suggested UTH is 645 mm, considering the 5th percentile of male SEH.

Designing a non-adjustable chair with a non-adjustable table set with a 100\% match for both males and females was impossible. Therefore, the dimensions were maintained to balance a reduced mismatch percentage for both males and females for the STH and UTH dimensions. For adjustable chairs with tables, it was possible to match the 100\% furniture dimensions for males and females. SH and STH showed better results for males and females because of the chair's height adjustability. SW, SD, and UTH showed results similar to those of non-adjustable chairs with table furniture sets. Although adjustable chairs showed improved results in some scenarios, we considered both types of furniture to be financially feasible and avoided design complexity. 

The proposed dimensions for both types of furniture followed international standards except for SD and SW. The SD and SW are slightly less than those of the European Committee for Standardization (2016)~\cite{EN1729-1}, but these values are very close to the standard values. The reason behind this is that the proposed dimensions are based on our collected samples that contained more male data and fewer female data. The HB is generally lower in males than in females.

VDT should be placed according to recent studies; computer monitor placement should be based on the needs of a particular user and should be performed within a modest height range below eye level with a head tilt that is physiologically beneficial \cite{woo_ergonomics_2016}. Occupational Safety and Health Administration (2008)~\cite{osha2008} suggests that monitors should be positioned 50–100 cm away from users while maintaining a viewing angle of $15^\circ$–$20^\circ$ \cite{weidling_vertical_2015}.

Using a keyboard and mouse is a frequent task in a computer lab, and it should be placed in the frequent or regular zone of the Barnes and Squires Work Envelope Model. This indicates that the keyboard and mouse should be placed within the 394 mm of table depth of 1194 mm and a table length \cite{norman_integrated_2013}.

The research is centered on the design of two sets of furniture identified within computer laboratories across various departments at KUET. Although the dimensions may exhibit variability, both types of furniture share similarities and are commonly employed in computer laboratories across diverse university settings. The proposed dimensions for these furniture sets were tailored to accommodate the anthropometric measurements of their users without introducing design intricacies or cost escalation.

The practical benefits of this research are significant for educational institutions, furniture manufacturers, and students. Inconsistencies in furniture dimensions were found by comprehensively analyzing anthropometric measurements and their correlation with furniture dimensions, which offers valuable insights into designing more comfortable and ergonomically sound furniture for university students. For educational institutions, this study suggests that investing in furniture tailored to students' body dimensions can contribute to a more conducive learning environment. Comfortable and ergonomic furniture can help students maintain better posture, reduce the risk of MSDs, and improve their overall well-being and academic performance. Furniture manufacturers can benefit from this research by using the proposed dimensions as a guideline for designing products that are better suited to the needs of students. By incorporating these dimensions into their designs, manufacturers can create furniture that is more comfortable, functional, and attractive for their target market.

Students stand to gain the most from this research as they are the end users of furniture. By using furniture designed with body dimensions in mind, students can enjoy a more comfortable and ergonomic learning experience. This can improve focus, concentration, and productivity and ultimately enhance academic performance.

This research also has managerial implications for educational institutions and stakeholders furnishing university spaces. Educational institutions can use these findings to inform strategic planning, prioritizing investments in ergonomic furniture tailored to students' anthropometric measurements. By incorporating the research findings into procurement requirements, institutions can ensure that they purchase furniture that meets students' ergonomic needs, thereby increasing their satisfaction and well-being. Based on this research, facility managers can optimize classroom and computer lab layouts, ensuring that furniture accommodates students' anthropometric measurements. This promotes a more comfortable and efficient learning environment and enhances student satisfaction, well-being, and academic performance. Educating students about the importance of ergonomic furniture and proper posture can prevent MSDs and related issues and improve student satisfaction, well-being, and academic performance. Establishing guidelines for furniture based on anthropometric criteria ensures compliance with relevant regulations and standards and provides students with safe, comfortable, and ergonomic learning environments. Furthermore, investing in ergonomic furniture tailored to anthropometric measurements can enhance health, comfort, productivity, and sustainable practices, leading to cost savings, an improved reputation, and legal compliance.

The limitations of this research include the reliance on a specific population (university students); the population size was 380. Although the number is statistically justified from a practical point of view, it should be increased further in future research. The exclusion of certain anthropometric measurements and the lack of consideration of other factors influencing furniture design (e.g., cultural preferences and environmental constraints) is another limitation of our research. Additionally, the proposed furniture dimensions may not be universally applicable and require further validation in other settings or populations. Future research could explore the implementation of the proposed dimensions in real-world classroom settings and assess their long-term impacts on student comfort and well-being using a larger population size. In addition, measuring the human body repeatedly would show slightly different results every time, which was neglected in our research. Further studies should investigate the effects of different furniture designs on student performance and productivity in academic settings.

\section{Conclusions}
\label{conclude}
This study underscores the critical importance of ergonomics in university computer laboratories, where students' prolonged interactions with computers necessitate optimal ergonomic conditions. By proposing anthropometric-based furniture dimensions specifically tailored for university students, we aim to enhance the ergonomic design of these environments. Significant disparities were identified through meticulous analysis of 11 key anthropometric measurements and comparison with existing furniture dimensions. Statistical validation through ANOVA reinforced the credibility of our proposed dimensions, showing reduced mismatch percentages compared to existing furniture. This indicates higher compatibility and potential improvements in student comfort, as well as a reduction in the risk of MSDs. Our study also revealed significant disparities between existing furniture dimensions and students' anthropometric data, particularly in crucial areas such as SW, BH, and STC. By identifying areas for improvement and proposing new dimensions aligned with students' anthropometric measurements and international standards, our research offers practical insights that can inform future furniture design efforts and contribute to developing ergonomic standards in educational institutions.

Implementing these proposed dimensions can substantially improve the ergonomic design of university computer laboratories, ultimately promoting student health, well-being, and academic success. Moreover, this study has significant practical implications for educational institutions, furniture manufacturers, and students alike. Investing in furniture tailored to students' body dimensions can create a more conducive learning environment, improving posture, reducing the risk of MSDs, and enhancing overall academic performance. By advocating for the adoption of ergonomically designed furniture, our research seeks to prioritize student welfare and productivity in university settings.

The contributions of the proposed work are as follows:
\begin{enumerate}
    \item This study highlights the importance of considering anthropometric measurements in furniture design for educational institutions, serving as a reference for future research and informing the design of ergonomic furniture in other educational settings.
    \item By identifying the mismatch between existing furniture dimensions and anthropometric measurements, this study offers practical insights for designing ergonomic furniture in university environments and for improving comfort, health, and productivity among students and staff.
    \item Designing ergonomic furniture that meets the specific needs of university students and staff may lead to potential cost savings for educational institutions through reduced absenteeism, increased productivity, and improved overall well-being.
    \item We collected and analyzed anthropometric data from university students to provide valuable insights into the body measurements of the target population.
    \item By comparing the mismatch between two different types of existing furniture and anthropometric data, this study offers a nuanced understanding of how different furniture designs affect users' comfort and well-being.
\end{enumerate}

This study also provides recommendations for optimized furniture design. The findings have significant practical implications, as they can guide the development of furniture that promotes better ergonomics and reduces the risk of MSDs among students. Moreover, this study's methodology and insights can be applied beyond the educational setting to inform furniture design for various industries and environments. This underscores the importance of considering cultural preferences and environmental constraints when designing furniture products. This study contributes to the growing body of ergonomics research and can potentially improve the well-being and productivity of individuals in diverse settings.

\section*{Ethical Approval}
The Office of the Director for Research \& Extension reviewed and approved this study under the approval code KUET/DRE/2023/15(6). Written informed consent was obtained from all participants prior to their inclusion in the study.

\section*{Data Availability Statement}
The dataset used in this research, entitled ``Anthropometric Data of KUET Students," is available for further exploration and can be accessed on Mendeley~\cite{jahin_anthropometric_2024}. It can be found at the following DOI link: \href{https://dx.doi.org/10.17632/kw7fd465v7.2}{https://dx.doi.org/10.17632/kw7fd465v7.2}.

\section*{Funding}
No financial support was provided by public, commercial, or non-profit funding bodies for this research.

\section*{CRediT authorship contribution statement}
\textbf{Anik Kumar Saha}: Conceptualization, Data curation, Formal analysis, Investigation, Methodology, Software, Writing - original draft, Visualization.
\textbf{Md Abrar Jahin}: Conceptualization, Data curation, Formal analysis, Investigation, Methodology, Software, Writing - original draft, Visualization.
\textbf{Md. Rafiquzzaman}: Supervision, Validation, Writing - review \& editing.
\textbf{M. F. Mridha}: Validation, Writing - review \& editing.

%%%% don't change the commands below %%%%%%%
%\twocolumn
% \bibliographystyle{apalike}
\bibliographystyle{elsarticle-num}
\bibliography{main}

\onecolumn
\newpage
\appendix

\section{NMQ Survey Questionnaire Format}
\begin{center}
Ergonomic Assessment
\end{center}
\begin{table*}[!ht]
\centering
\begin{tabular}{| l |}
\hline
Serial no: \hspace{10cm}
\\
Age: \hspace{10cm}
\\
Gender: \hspace{10cm}
\\
Date of Assessment: \hspace{9cm}
\\
\hline
\end{tabular}
\end{table*}
Question: Do you experience pain or discomfort in any of the following body parts when using your computer? If “yes,” please mention which part of your body.\\
\\
\begin{enumerate*}[label=(\alph*)]
    \item Neck 
    \item Shoulders/upper arm 
    \item Upper back 
    \item Lower back 
    \item Elbow/forearm 
    \item Wrist/Hand 
    \item Hips/Buttocks/Thighs 
    \item Knees and legs 
    \item Feet/Ankles 
\end{enumerate*}\\
\\
Please mention in which form is affected.
\begin{enumerate}[label=(\roman*)]
\item Constantly (most time of the day) 
\item Frequently (more than four times a month) 
\item Occasionally (two to four times a month)
\item Never
\end{enumerate}

\newpage
\section{Supplementary Tables and Figures}
\setcounter{table}{0}
\renewcommand{\thetable}{A\arabic{table}}

\begin{table*}[!ht]
\centering
\caption{Descriptive statistics of the anthropometric measurements taken for both male and female participants (in mm)}
\label{tab:4.1}
\resizebox{\linewidth}{!}{%
\begin{tabular}{|c|c|c|c|c|c|c|c|c|} 
\hline
\multirow{2}{*}{\textbf{Measurements}} & \multirow{2}{*}{\textbf{Gender}} & \multirow{2}{*}{\textbf{Min}} & \multirow{2}{*}{\textbf{Max}} & \multirow{2}{*}{\textbf{Mean}} & \multirow{2}{*}{\textbf{Standard Deviation}} & \multicolumn{3}{c|}{\textbf{Percentile}} \\ 
\cline{7-9}
 &  &  &  &  &  & \textbf{5\textsuperscript{th}~} & \textbf{50\textsuperscript{th}~} & \textbf{95\textsuperscript{th}~} \\ 
\hline
\multirow{2}{*}{PH} & Male & 414 & 489 & 444 & 13 & 423 & 445 & 466 \\ 
\cline{2-9}
 & Female & 373 & 447 & 415 & 14 & 395 & 414 & 438 \\ 
\hline
\multirow{2}{*}{SEH} & Male & 188 & 308 & 234 & 17 & 205 & 235 & 261 \\ 
\cline{2-9}
 & Female & 191 & 282 & 231 & 17 & 200 & 230 & 257 \\ 
\hline
\multirow{2}{*}{BPL} & Male & 402 & 498 & 454 & 18 & 422 & 454 & 483 \\ 
\cline{2-9}
 & Female & 394 & 478 & 447 & 17 & 412 & 448 & 471 \\ 
\hline
\multirow{2}{*}{AL} & Male & 331 & 394 & 364 & 11 & 346 & 364 & 382 \\ 
\cline{2-9}
 & Female & 322 & 366 & 342 & 10 & 327 & 342 & 356 \\ 
\hline
\multirow{2}{*}{TT} & Male & 131 & 178 & 156 & 8 & 143 & 156 & 169 \\ 
\cline{2-9}
 & Female & 126 & 175 & 145 & 10 & 129 & 145 & 161 \\ 
\hline
\multirow{2}{*}{SSH} & Male & 470 & 556 & 512 & 15 & 489 & 512 & 539 \\ 
\cline{2-9}
 & Female & 447 & 522 & 488 & 13 & 467 & 489 & 509 \\ 
\hline
\multirow{2}{*}{SCH} & Male & 484 & 540 & 512 & 10 & 495 & 511 & 528 \\ 
\cline{2-9}
 & Female & 469 & 524 & 494 & 11 & 475 & 494 & 510 \\ 
\hline
\multirow{2}{*}{EFL} & Male & 426 & 471 & 450 & 8 & 437 & 450 & 462 \\ 
\cline{2-9}
 & Female & 385 & 428 & 407 & 10 & 392 & 408 & 421 \\ 
\hline
\multirow{2}{*}{SEB} & Male & 396 & 492 & 447 & 15 & 424 & 448 & 470 \\ 
\cline{2-9}
 & Female & 390 & 464 & 423 & 17 & 396 & 421 & 453 \\ 
\hline
\multirow{2}{*}{HB} & Male & 328 & 382 & 351 & 9 & 335 & 350 & 365 \\ 
\cline{2-9}
 & Female & 344 & 390 & 366 & 8 & 353 & 367 & 379 \\ 
\hline
\multirow{2}{*}{BKL} & Male & 496 & 548 & 521 & 10 & 505 & 522 & 536 \\ 
\cline{2-9}
 & Female & 475 & 546 & 509 & 14 & 489 & 509 & 530 \\
\hline
\end{tabular}
}
\end{table*}

\setcounter{figure}{0}

\begin{figure*}[!ht]
\centering
    \begin{subfigure}{0.85\linewidth}
        \includegraphics[width=\linewidth]{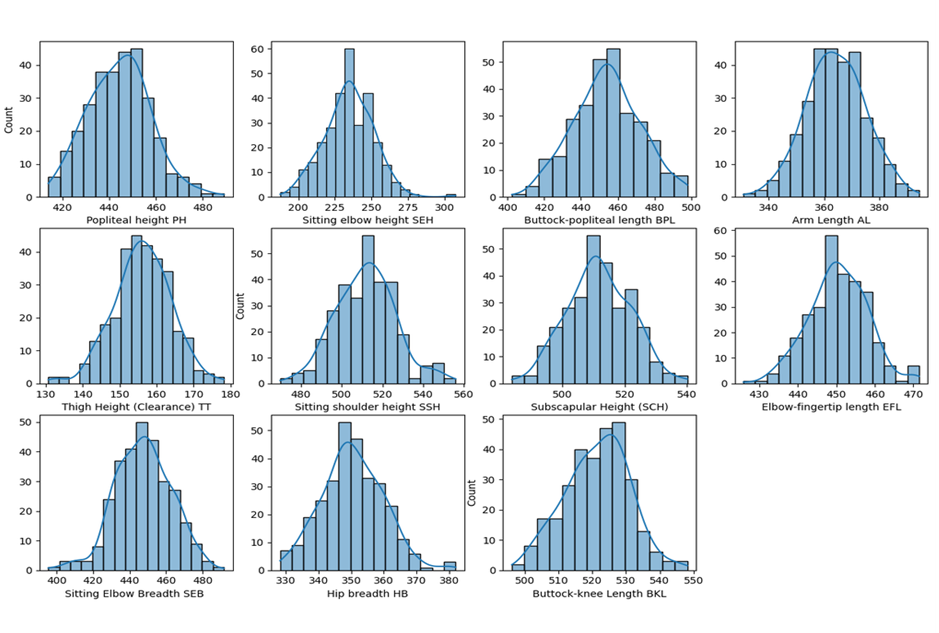}
        \caption{Distribution of male anthropometric data}
        \label{fig:4.1}
    \end{subfigure}
    \begin{subfigure}{0.85\linewidth}
        \centering
        \includegraphics[width=\linewidth]{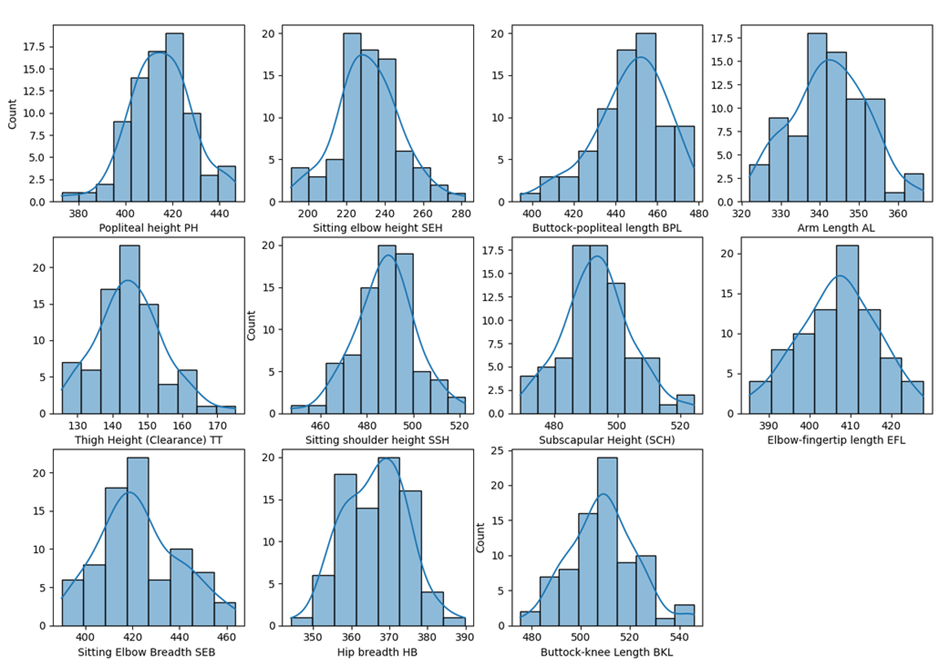}
        \caption{Distribution of female anthropometric data}
        \label{fig:4.2}
    \end{subfigure}
\caption{Anthropometric data showing normality (a bell-shaped curve) in the case of both (a) male and (b) female distributions.}
\end{figure*}

\begin{table*}[!ht]
\centering
\caption{ANOVA test results for non-adjustable furniture dimensions}
\label{tab:4.5}
\resizebox{\linewidth}{!}{%
\begin{tabular}{|c|l|l|l|l|l|l|l|l|} 
\hline
\multirow{2}{*}{\textbf{Furniture Dimensions vs Anthropometric Dimensions}} & \multirow{2}{*}{\textbf{Gender}} & \multirow{2}{*}{\textbf{Value Type}} & \multicolumn{3}{l|}{\textbf{Percentile}} & \multirow{2}{*}{\textbf{F -value}} & \multirow{2}{*}{\textbf{p - value}} & \multirow{2}{*}{\textbf{Decision}} \\ 
\cline{4-6}
 &  &  & \textbf{5th~} & \textbf{50th~} & \textbf{95th~} &  &  &  \\ 
\hline
\multirow{4}{*}{SH vs PH} & \multirow{2}{*}{Male} & Observed & 423 & 445 & 466 & \multirow{2}{*}{0.71} & \multirow{2}{*}{0.447} & \multirow{2}{*}{Accept} \\ 
\cline{3-6}
 &  & Expected & 444 & 457 & 471 &  &  &  \\ 
\cline{2-9}
 & \multirow{2}{*}{Female} & Observed & 395 & 415 & 438 & \multirow{2}{*}{7.83} & \multirow{2}{*}{0.049} & \multirow{2}{*}{Reject} \\ 
\cline{3-6}
 &  & Expected & 444 & 457 & 471 &  &  &  \\ 
\hline
\multirow{4}{*}{SW vs HB} & \multirow{2}{*}{Male} & Observed & 335 & 350 & 365 & \multirow{2}{*}{11.19} & \multirow{2}{*}{0.029} & \multirow{2}{*}{Reject} \\ 
\cline{3-6}
 &  & Expected & 377 & 394 & 410 &  &  &  \\ 
\cline{2-9}
 & \multirow{2}{*}{Female} & Observed & 353 & 367 & 379 & \multirow{2}{*}{4.51} & \multirow{2}{*}{0.101} & \multirow{2}{*}{Accept} \\ 
\cline{3-6}
 &  & Expected & 376 & 394 & 411 &  &  &  \\ 
\hline
\multirow{4}{*}{SD vs BPL} & \multirow{2}{*}{Male} & Observed & 423 & 454 & 484 & \multirow{2}{*}{5.38} & \multirow{2}{*}{0.081} & \multirow{2}{*}{Accept} \\ 
\cline{3-6}
 &  & Expected & 389 & 406 & 424 &  &  &  \\ 
\cline{2-9}
 & \multirow{2}{*}{Female} & Observed & 413 & 448 & 471 & \multirow{2}{*}{3.61} & \multirow{2}{*}{0.13} & \multirow{2}{*}{Accept} \\ 
\cline{3-6}
 &  & Expected & 389 & 406 & 424 &  &  &  \\ 
\hline
\multirow{4}{*}{BH vs SSH} & \multirow{2}{*}{Male} & Observed & 489 & 512 & 539 & \multirow{2}{*}{173.83} & \multirow{2}{*}{0} & \multirow{2}{*}{Reject} \\ 
\cline{3-6}
 &  & Expected & 295 & 305 & 315 &  &  &  \\ 
\cline{2-9}
 & \multirow{2}{*}{Female} & Observed & 467 & 489 & 509 & \multirow{2}{*}{184.73} & \multirow{2}{*}{0} & \multirow{2}{*}{Reject} \\ 
\cline{3-6}
 &  & Expected & 295 & 305 & 315 &  &  &  \\ 
\hline
\multirow{4}{*}{BW vs HB} & \multirow{2}{*}{Male} & Observed & 335 & 350 & 365 & \multirow{2}{*}{0.28} & \multirow{2}{*}{0.625} & \multirow{2}{*}{Accept} \\ 
\cline{3-6}
 &  & Expected & 347 & 356 & 364 &  &  &  \\ 
\cline{2-9}
 & \multirow{2}{*}{Female} & Observed & 353 & 367 & 379 & \multirow{2}{*}{1.48} & \multirow{2}{*}{0.29} & \multirow{2}{*}{Accept~} \\ 
\cline{3-6}
 &  & Expected & 347 & 356 & 364 &  &  &  \\ 
\hline
\multirow{4}{*}{UEB vs SCH} & \multirow{2}{*}{Male} & Observed & 495 & 511 & 528 & \multirow{2}{*}{70.93} & \multirow{2}{*}{0.001} & \multirow{2}{*}{Reject} \\ 
\cline{3-6}
 &  & Expected & 392 & 406 & 420 &  &  &  \\ 
\cline{2-9}
 & \multirow{2}{*}{Female} & Observed & 475 & 494 & 510 & \multirow{2}{*}{44.56} & \multirow{2}{*}{0.003} & \multirow{2}{*}{Reject} \\ 
\cline{3-6}
 &  & Expected & 392 & 406 & 420 &  &  &  \\ 
\hline
\multirow{4}{*}{STH vs SEH} & \multirow{2}{*}{Male} & Observed & 206 & 235 & 261 & \multirow{2}{*}{0.17} & \multirow{2}{*}{0.694} & \multirow{2}{*}{Accept} \\ 
\cline{3-6}
 &  & Expected & 231 & 241 & 252 &  &  &  \\ 
\cline{2-9}
 & \multirow{2}{*}{Female} & Observed & 201 & 230 & 258 & \multirow{2}{*}{0.45} & \multirow{2}{*}{0.538} & \multirow{2}{*}{Accept} \\ 
\cline{3-6}
 &  & Expected & 231 & 241 & 252 &  &  &  \\ 
\hline
\multirow{4}{*}{STC vs TT} & \multirow{2}{*}{Male} & Observed & 143 & 156 & 169 & \multirow{2}{*}{58.13} & \multirow{2}{*}{0.002} & \multirow{2}{*}{Reject} \\ 
\cline{3-6}
 &  & Expected & 80 & 89 & 97 &  &  &  \\ 
\cline{2-9}
 & \multirow{2}{*}{Female} & Observed & 129 & 145 & 161 & \multirow{2}{*}{28.58} & \multirow{2}{*}{0.006} & \multirow{2}{*}{Reject} \\ 
\cline{3-6}
 &  & Expected & 80 & 89 & 97 &  &  &  \\ 
\hline
\multirow{4}{*}{TL vs BKL} & \multirow{2}{*}{Male} & Observed & 505 & 522 & 536 & \multirow{2}{*}{13.8} & \multirow{2}{*}{0.021} & \multirow{2}{*}{Reject} \\ 
\cline{3-6}
 &  & Expected & 474 & 483 & 492 &  &  &  \\ 
\cline{2-9}
 & \multirow{2}{*}{Female} & Observed & 489 & 509 & 530 & \multirow{2}{*}{4.37} & \multirow{2}{*}{0.105} & \multirow{2}{*}{Accept} \\ 
\cline{3-6}
 &  & Expected & 474 & 483 & 492 &  &  &  \\
\hline
\end{tabular}
}
\end{table*}

\begin{table*}[!ht]
\centering
\caption{ANOVA test results for adjustable chair with non-adjustable table dimensions}
\label{tab:4.6}
\resizebox{\linewidth}{!}{%
\begin{tabular}{|c|l|l|l|l|l|l|l|l|} 
\hline
\multirow{2}{*}{\textbf{Furniture Dimensions vs Anthropometric Dimensions}} & \multirow{2}{*}{\textbf{Gender}} & \multirow{2}{*}{\textbf{Value Type}} & \multicolumn{3}{l|}{\textbf{Percentile}} & \multirow{2}{*}{\textbf{F-value}} & \multirow{2}{*}{\textbf{p-value}} & \multirow{2}{*}{\textbf{Decision}} \\ 
\cline{4-6}
 &  &  & \textbf{5th~} & \textbf{50th~} & \textbf{95th~} &  &  &  \\ 
\hline
\multirow{4}{*}{SH (lowest limit) vs PH} & \multirow{2}{*}{Male} & Observed & 423 & 445 & 466 & \multirow{2}{*}{0.527} & \multirow{2}{*}{0.508} & \multirow{2}{*}{Accept} \\ 
\cline{3-6}
 &  & Expected & 410 & 432 & 453 &  &  &  \\ 
\cline{2-9}
 & \multirow{2}{*}{Female} & Observed & 395 & 415 & 438 & \multirow{2}{*}{0.796} & \multirow{2}{*}{0.423} & \multirow{2}{*}{Accept} \\ 
\cline{3-6}
 &  & Expected & 410 & 432 & 453 &  &  &  \\ 
\hline
\multirow{4}{*}{SH (highest limit) vs PH} & \multirow{2}{*}{Male} & Observed & 423 & 445 & 466 & \multirow{2}{*}{19.804} & \multirow{2}{*}{0.011} & \multirow{2}{*}{Reject} \\ 
\cline{3-6}
 &  & Expected & 507 & 533 & 560 &  &  &  \\ 
\cline{2-9}
 & \multirow{2}{*}{Female} & Observed & 395 & 415 & 438 & \multirow{2}{*}{35.008} & \multirow{2}{*}{0.004} & \multirow{2}{*}{Reject} \\ 
\cline{3-6}
 &  & Expected & 507 & 533 & 560 &  &  &  \\ 
\hline
\multirow{4}{*}{SW vs HB} & \multirow{2}{*}{Male} & Observed & 335 & 350 & 365 & \multirow{2}{*}{45.836} & \multirow{2}{*}{0.002} & \multirow{2}{*}{~Reject} \\ 
\cline{3-6}
 &  & Expected & 434 & 457 & 480 &  &  &  \\ 
\cline{2-9}
 & \multirow{2}{*}{Female} & Observed & 353 & 367 & 379 & \multirow{2}{*}{35.436} & \multirow{2}{*}{0.004} & \multirow{2}{*}{Reject} \\ 
\cline{3-6}
 &  & Expected & 434 & 457 & 480 &  &  &  \\ 
\hline
\multirow{4}{*}{SD vs BPL} & \multirow{2}{*}{Male} & Observed & 423 & 454 & 484 & \multirow{2}{*}{1.023} & \multirow{2}{*}{0.369} & \multirow{2}{*}{Accept} \\ 
\cline{3-6}
 &  & Expected & 410 & 432 & 453 &  &  &  \\ 
\cline{2-9}
 & \multirow{2}{*}{Female} & Observed & 413 & 448 & 471 & \multirow{2}{*}{0.337} & \multirow{2}{*}{0.593} & \multirow{2}{*}{Accept} \\ 
\cline{3-6}
 &  & Expected & 410 & 432 & 453 &  &  &  \\ 
\hline
\multirow{4}{*}{BH vs SSH} & \multirow{2}{*}{Male} & Observed & 489 & 512 & 539 & \multirow{2}{*}{148.401} & \multirow{2}{*}{0} & \multirow{2}{*}{Reject} \\ 
\cline{3-6}
 &  & Expected & 290 & 305 & 320 &  &  &  \\ 
\cline{2-9}
 & \multirow{2}{*}{Female} & Observed & 467 & 489 & 509 & \multirow{2}{*}{149.574} & \multirow{2}{*}{0} & \multirow{2}{*}{Reject} \\ 
\cline{3-6}
 &  & Expected & 290 & 305 & 320 &  &  &  \\ 
\hline
\multirow{4}{*}{BW vs HB} & \multirow{2}{*}{Male} & Observed & 335 & 350 & 365 & \multirow{2}{*}{9.223} & \multirow{2}{*}{0.039} & \multirow{2}{*}{Reject} \\ 
\cline{3-6}
 &  & Expected & 374 & 394 & 413 &  &  &  \\ 
\cline{2-9}
 & \multirow{2}{*}{Female} & Observed & 353 & 367 & 379 & \multirow{2}{*}{3.926} & \multirow{2}{*}{0.119} & \multirow{2}{*}{Accept} \\ 
\cline{3-6}
 &  & Expected & 374 & 394 & 413 &  &  &  \\ 
\hline
\multirow{4}{*}{UEB vs SCH} & \multirow{2}{*}{Male} & Observed & 495 & 511 & 528 & \multirow{2}{*}{48.612} & \multirow{2}{*}{0.002} & \multirow{2}{*}{Reject} \\ 
\cline{3-6}
 &  & Expected & 386 & 406 & 427 &  &  &  \\ 
\cline{2-9}
 & \multirow{2}{*}{Female} & Observed & 475 & 494 & 510 & \multirow{2}{*}{31.274} & \multirow{2}{*}{0.005} & \multirow{2}{*}{Reject} \\ 
\cline{3-6}
 &  & Expected & 386 & 406 & 427 &  &  &  \\ 
\hline
\multirow{4}{*}{STH (lowest limit) vs SEH} & \multirow{2}{*}{Male} & Observed & 206 & 235 & 261 & \multirow{2}{*}{0.101} & \multirow{2}{*}{0.766} & \multirow{2}{*}{Accept} \\ 
\cline{3-6}
 &  & Expected & 217 & 229 & 240 &  &  &  \\ 
\cline{2-9}
 & \multirow{2}{*}{Female} & Observed & 201 & 230 & 258 & \multirow{2}{*}{0.003} & \multirow{2}{*}{0.962} & \multirow{2}{*}{Accept} \\ 
\cline{3-6}
 &  & Expected & 217 & 229 & 240 &  &  &  \\ 
\hline
\multirow{4}{*}{STH (highest limit) vs SEH} & \multirow{2}{*}{Male} & Observed & 206 & 235 & 261 & \multirow{2}{*}{26.745} & \multirow{2}{*}{0.007} & \multirow{2}{*}{Reject} \\ 
\cline{3-6}
 &  & Expected & 314 & 330 & 347 &  &  &  \\ 
\cline{2-9}
 & \multirow{2}{*}{Female} & Observed & 201 & 230 & 258 & \multirow{2}{*}{28.057} & \multirow{2}{*}{0.006} & \multirow{2}{*}{Reject} \\ 
\cline{3-6}
 &  & Expected & 314 & 330 & 347 &  &  &  \\ 
\hline
\multirow{4}{*}{STC (lowest limit) vs TT} & \multirow{2}{*}{Male} & Observed & 143 & 156 & 169 & \multirow{2}{*}{61.03} & \multirow{2}{*}{0.001} & \multirow{2}{*}{Reject} \\ 
\cline{3-6}
 &  & Expected & 90 & 95 & 100 &  &  &  \\ 
\cline{2-9}
 & \multirow{2}{*}{Female} & Observed & 129 & 145 & 161 & \multirow{2}{*}{26.571} & \multirow{2}{*}{0.007} & \multirow{2}{*}{Reject} \\ 
\cline{3-6}
 &  & Expected & 90 & 95 & 100 &  &  &  \\ 
\hline
\multirow{4}{*}{STC (highest limit) vs TT} & \multirow{2}{*}{Male} & Observed & 143 & 156 & 169 & \multirow{2}{*}{19.526} & \multirow{2}{*}{0.012} & \multirow{2}{*}{~Reject} \\ 
\cline{3-6}
 &  & Expected & 187 & 197 & 207 &  &  &  \\ 
\cline{2-9}
 & \multirow{2}{*}{Female} & Observed & 129 & 145 & 161 & \multirow{2}{*}{22.616} & \multirow{2}{*}{0.009} & \multirow{2}{*}{Reject} \\ 
\cline{3-6}
 &  & Expected & 187 & 197 & 207 &  &  &  \\ 
\hline
\multirow{4}{*}{TL vs BKL} & \multirow{2}{*}{Male} & Observed & 505 & 522 & 536 & \multirow{2}{*}{16.011} & \multirow{2}{*}{0.016} & \multirow{2}{*}{Reject} \\ 
\cline{3-6}
 &  & Expected & 434 & 457 & 480 &  &  &  \\ 
\cline{2-9}
 & \multirow{2}{*}{Female} & Observed & 489 & 509 & 530 & \multirow{2}{*}{8.739} & \multirow{2}{*}{0.042} & \multirow{2}{*}{Reject~} \\ 
\cline{3-6}
 &  & Expected & 434 & 457 & 480 &  &  &  \\
\hline
\end{tabular}
}
\end{table*}

\end{document}